\documentclass[a4paper]{jpconf}
\usepackage{graphicx}

\usepackage{citesort}

\usepackage{amssymb}

\begin{document}

\nocite{*} 

\title{Cosmology with 21cm intensity mapping}

\author{I P Carucci $^{1, 2, 3}$}

\address{$^1$ SISSA - International School for Advanced Studies, Via Bonomea 265, 34136 Trieste, Italy}
\address{$^2$ INFN sez. Trieste, Via Valerio 2, 34127, Trieste, Italy}
\address{$^3$ LUTH, Observatoire de Meudon, 5 Place Janssen, 92195 Meudon, France}

\ead{ipcarucci@sissa.it}

\begin{abstract}
The nature of the most abundant components of the Universe, dark energy and dark matter, is still to be uncovered. I  tackle this subject considering a novel cosmological probe: the neutral hydrogen emitted  21cm radiation, observed with the intensity mapping technique. I analyse competitive and realistic dark energy and dark matter models and show how they produce distinctive and detectable effects on the 21cm signal. Moreover, I provide radio telescope forecasts showing how these models will be distinguishable in an  unprecedented way.
\end{abstract}

\section{Introduction}

The current standard  cosmological model, $\Lambda$CDM,  has outstandingly provided the theoretical framework to interpret a wide variety of observables \cite{Planck_2015}, however this success is linked to the existence of two dark components. The Universe is now experiencing an accelerated expansion generated by an unknown Dark Energy (DE) component \cite{riess_1998,Perlmutter_1999}, while an invisible Dark Matter (DM) drives the formation of all visible structures (see e.g. \cite{tegmark_2004,clowe_2006,massey_2007}). Effects of both components are embedded in the spatial distribution of matter in the Universe, that we can trace thanks to different observables such as galaxy clustering, weak lensing, abundance of galaxy clusters and  Lyman-$\alpha$ forest.

In the upcoming years, observations of the redshifted 21cm radiation from neutral hydrogen (HI) will allow us to map the large scale structure of the Universe with an unprecedented precision on redshifts not accessible with the above probes  \cite{Bull_2014,Camera_2013}. The idea is to carry out Intensity Mapping (IM) observations: measuring with low angular resolution  the integrated 21cm emission coming  from unresolved sources contained in large patches of the sky. Given its spectroscopic nature and the large volumes sampled, with 21cm IM we will soon start doing cosmology. Among current and upcoming instruments able to employ this technique we find: CHIME\footnote{http://chime.phas.ubc.ca/}, BINGO\footnote{http://www.jb.man.ac.uk/research/BINGO/}, ORT\footnote{http://rac.ncra.tifr.res.in/}, FAST\footnote{http://fast.bao.ac.cn/en/} and ultimately SKA\footnote{https://www.skatelescope.org/} and its pathfinders.

In this paper, I will illustrate the imprint on the 21cm IM signal power spectrum left by non-standard cosmological models otherwise indistinguishable from $\Lambda$CDM, quantifying these effects in the context of realistic radio-telescope forecasts. The work presented is based on \cite{carucci_2015,carucci_2017}.

\section{Modelling the HI spatial distribution}

Radio telescopes detect the redshifted 21cm radiation emitted by neutral hydrogen, thus we need to model the distribution of HI in the Universe. Here I summarise how we carry out the modelling and the analysis. After cosmic reionization, it is believed that the quasi totality of HI resides in DM halos, that are most accurately modelled with numerical simulations being halos biased tracers of the DM density field.  We make use of simulations that we present together with the cosmologies considered. After having identified the halos (in all cases with a FoF algorithm \cite{FoF}), we proceed at populating them with HI following the prescriptions presented in \cite{carucci_2015,carucci_2017}. Once the gas is distributed, computing the expected 21cm signal is straightforward. To better assess the significance of the computed signals, we forecast errors with which SKA1-LOW and SKA1-MID\footnote{The choice of the telescope depends on the considered redshift of observation.}  will measure them, taking care of the instrumental noise only, i.e. assuming that astrophysical foregrounds, radio frequency interference and others nuisances have been already removed from the observed data. 

\section{Non-standard dark matter models}

\begin{figure}[t]
	\begin{minipage}{0.60\textwidth}
		\includegraphics[width=\textwidth]{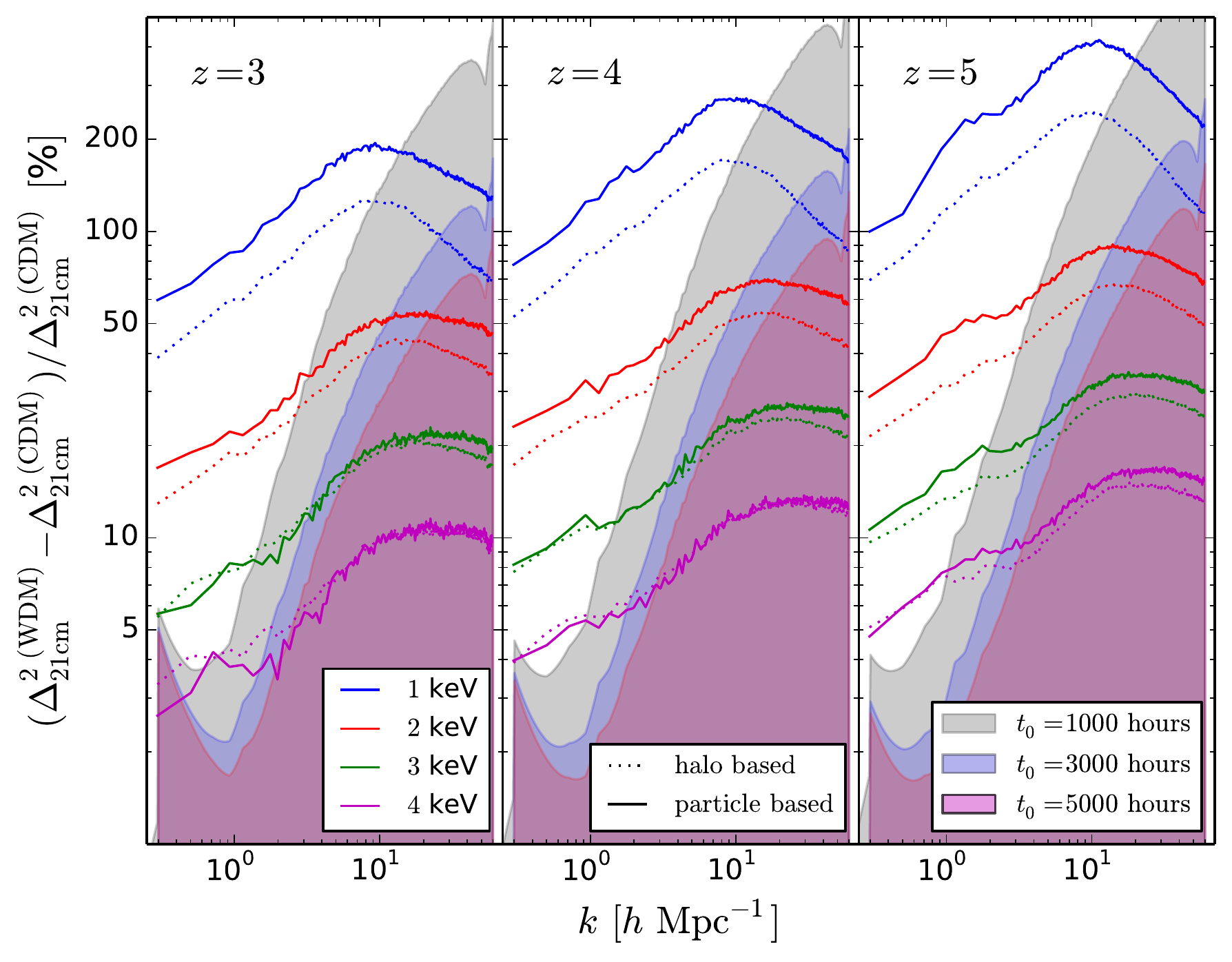}
	\end{minipage}\hspace{0.04\textwidth}%
	\begin{minipage}{0.36\textwidth}
		\caption{\label{WDM}21cm intensity map power spectrum of the thermal  WDM models relative to the reference $\Lambda$CDM predictions at  $z=3$ (left), $z=4$ (middle) and $z=5$ (right), dotted and solid lines refer to different HI distribution methods (see \cite{carucci_2015} for details). The shaded area represents the expected errors from SKA1-LOW measurements for the reference $\Lambda$CDM model,  $\sigma[P^{\Lambda{\rm CDM}}_{21{\rm cm}}(k)] / P^{\Lambda{\rm CDM}}_{21{\rm cm}}(k)$, assuming $t_0 = 1000$ (grey), $t_0 = 3000$ (blue) and $t_0 = 5000$ (fuchsia) observing hours.}
	\end{minipage}
\end{figure}

\begin{figure}[t]
	\begin{minipage}{0.60\textwidth}
		\includegraphics[width=\textwidth]{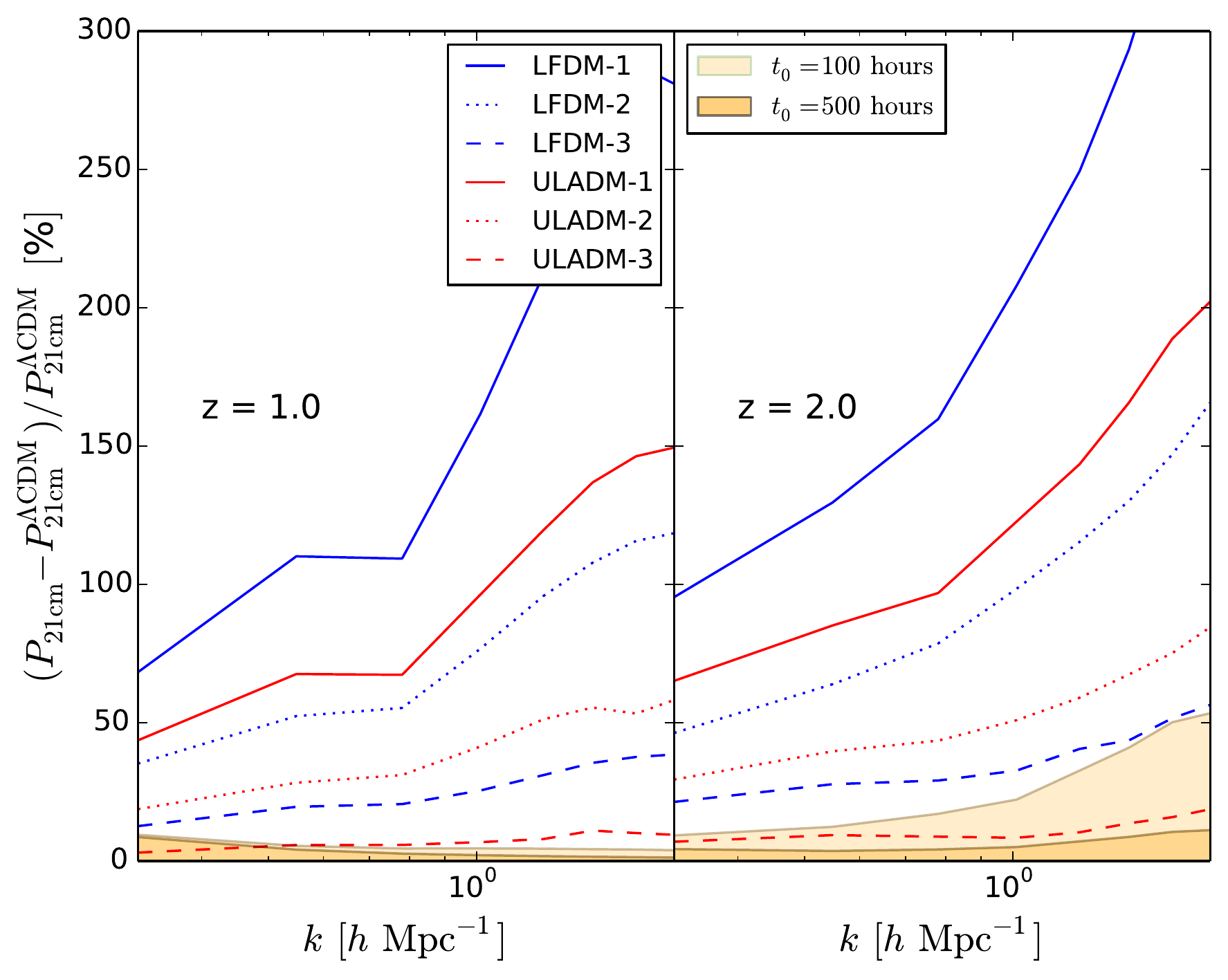}
	\end{minipage}\hspace{0.04\textwidth}%
	\begin{minipage}{0.36\textwidth}
\caption{\label{SKA_dm} 21cm intensity map power spectrum of the LFDM and ULADM models relative to the reference $\Lambda$CDM-S predictions at $z=1$ (left panel) and $2$ (right panel) respectively. The shaded area represents the expected errors from SKA1-MID measurements for the reference $\Lambda$CDM-S model,  $\sigma[P^{\Lambda{\rm CDM}}_{21{\rm cm}}(k)] / P^{\Lambda{\rm CDM}}_{21{\rm cm}}(k)$, assuming $t_0 = 100$ (light shaded area) and $t_0 = 500$ (dark shaded area) observing hours. Figure taken from \cite{carucci_2017}.}
	\end{minipage}
\end{figure}

The $\Lambda$CDM model is constituted by a perfectly collisionless \emph{cold} dark matter component (CDM), i.e. with negligible thermal velocities on all scales at high redshift. However, tensions exist at small scales between CDM predictions and various observational probes (see \cite{weinberg_2015} for a review). There are various theoretical DM models that can alleviate these tensions while preserving the success of CDM on large scales, i.e. they are indistinguishable from CDM at large scales. They do so by suppressing gravitational clustering at small scales, i.e. DM can not significantly cluster below a specific length. The overall effect is to retain a matter power spectrum with a cut-off on a characteristic scale that depends on the nature of the DM particle. Here we take into consideration three families of models:
\begin{enumerate}
	\item WDM: thermal warm DM, with intrinsic momenta derived from a Fermi-Dirac distribution;
	\item ULADM: ultra-light axion DM models, in the literature sometime called \emph{fuzzy} DM \cite{Marsh2016};
	\item LFDM: late-forming DM models \cite{Das2011,Agarwal2015a}.
\end{enumerate}

All these models are characterised by the above invoked suppression of the amplitude of matter density fluctuations at small scales below a characteristic length that for WDM models depends on the thermal relic mass, for ULADM on the axion particle mass, while in the case of LFDM models depends on the phase transition redshift. 

In the case of WDM, we consider a set of five high-resolution hydrodynamic N-body simulations presented in \cite{carucci_2015}. They follow the evolution of $512^3$ CDM/WDM and $512^3$ baryon (gas+stars) particles within simulation boxes of comoving sizes equal to 30 $h^{-1}{\rm Mpc}$. 
We have simulated five different cosmological models: one model with CDM and four models with WDM, each of them with a different Fermi-Dirac mass of the WDM particles: 1 keV, 2 keV, 3 keV and 4 keV. This keV range is that of interest since today tightest constraints come from the Lyman-$\alpha$ forest high redshift power spectrum  and point towards lower limits of $m_{\rm WDM}\gtrsim3.5 - 5.3~{\rm keV}$ (at 2$\sigma$ C.L.) \cite{irsic_2017a}. The values of the cosmological parameters are common to all five simulations: $\Omega_{\rm m}=0.3175$, $\Omega_{\rm b}=0.049$, $\Omega_\Lambda=0.6825$, $h=0.6711$, $n_s=0.9624$ and $\sigma_8=0.834$. We analyse snapshots at redshifts $z=1$, $2$ and $3$.

As shown in \cite{carucci_2015}, for WDM masses between 3 and 4 keV we observe a suppression in power up to $\sim10\%$ on very small scales and a reduction in the number of $10^9~h^{-1}{\rm M}_{\odot}$ halos of the order of 20-40\% compared to the CDM case. This lack of low mass halos, results in a higher amplitude of the 21cm power spectrum for the WDM than CDM, because the spatial distribution of neutral hydrogen is more strongly clustered in the models with WDM. In Fig.~\ref{WDM} we show the relative difference in the signal power spectrum between the models with WDM and the model with CDM. The uncertainties with which the SKA1-LOW radio-telescope will measure the 21cm power spectrum are highlighted with shaded regions. We find that with a reasonable observational time of $t_0=1000$ hours the WDM models with 1, 2, and 3 keV can be distinguished from  CDM. On the other hand, the predicted power spectrum of the 4 keV WDM model is consistent with the one predicted by CDM at $\sim 1\sigma$ confidence level at $z=5$. 

For the last two families of non-standard DM, ultra-light axions and late forming models, we consider N-body simulations presented in \cite{Corasaniti2017} with box of comoving size of $27.5$ Mpc $h^{-1}$, that follow $1024^3$ particles. They consist of three ULADM models\footnote{
	Note that the ULADM models investigated  are in disagreement with the recent  constraints obtained at $z>3$ from
	the Lyman-$\alpha$ forest  which result in a lower limit at the 2$\sigma$ level of $\sim 2 \times 10^{-21}$ eV \cite{irsic_2017b}. 
	}
: with axion mass of $m_a=1.56\times 10^{-22}$ eV (ULADM-1), $4.16\times 10^{-22}$ eV (ULADM-2) and $1.54\times 10^{-21}$ eV (ULADM-3), and of three late-forming DM models with transition redshift $z_t=5\times 10^5$ (LFDM-1), $8\times 10^5$ (LFDM-2) and $15\times 10^{15}$ (LFDM-3). The cosmological model parameters have been set to those of a reference $\Lambda$CDM simulation ($\Lambda$CDM-S) of the same box size  and with equal number of particles with $\Omega_m=0.3$, $\Omega_b=0.046$, $h=0.7$, $\sigma_8=0.8$ and $n_s=0.99$. We analyse snapshots at $z=1$ and $2$.

As for WDM, also these models are characterised by a suppression in the number of low mass halos. In Fig.~\ref{SKA_dm} we plot the 21cm IM spectra in the case of the ULADM and LFDM models relative to the reference $\Lambda$CDM-S. Again, we address significance by showing the forecasted error for the SKA1-MID radio telescope in interferometry for the reference $\Lambda$CDM-S model 21cm IM measurement. 

\medskip
The 21cm power spectra of the non-standard DM cosmologies analysed do not display any small scale cut-off as the matter power spectra do and this is consequence of neutral hydrogen being hosted in DM halos: these scenarios suppress halos abundance and force  HI to be more clustered in the  halos that are left (that are also the more massive, i.e. more biased). This results in expecting a higher 21cm IM signal in these scenarios, enough higher  than instrumental noise to tell them apart from a standard CDM paradigm (as shown in Figs.~\ref{WDM}-\ref{SKA_dm}).

\section{Dynamical dark energy}

\begin{figure}[t]
	\begin{minipage}{0.60\textwidth}
		\includegraphics[width=\textwidth]{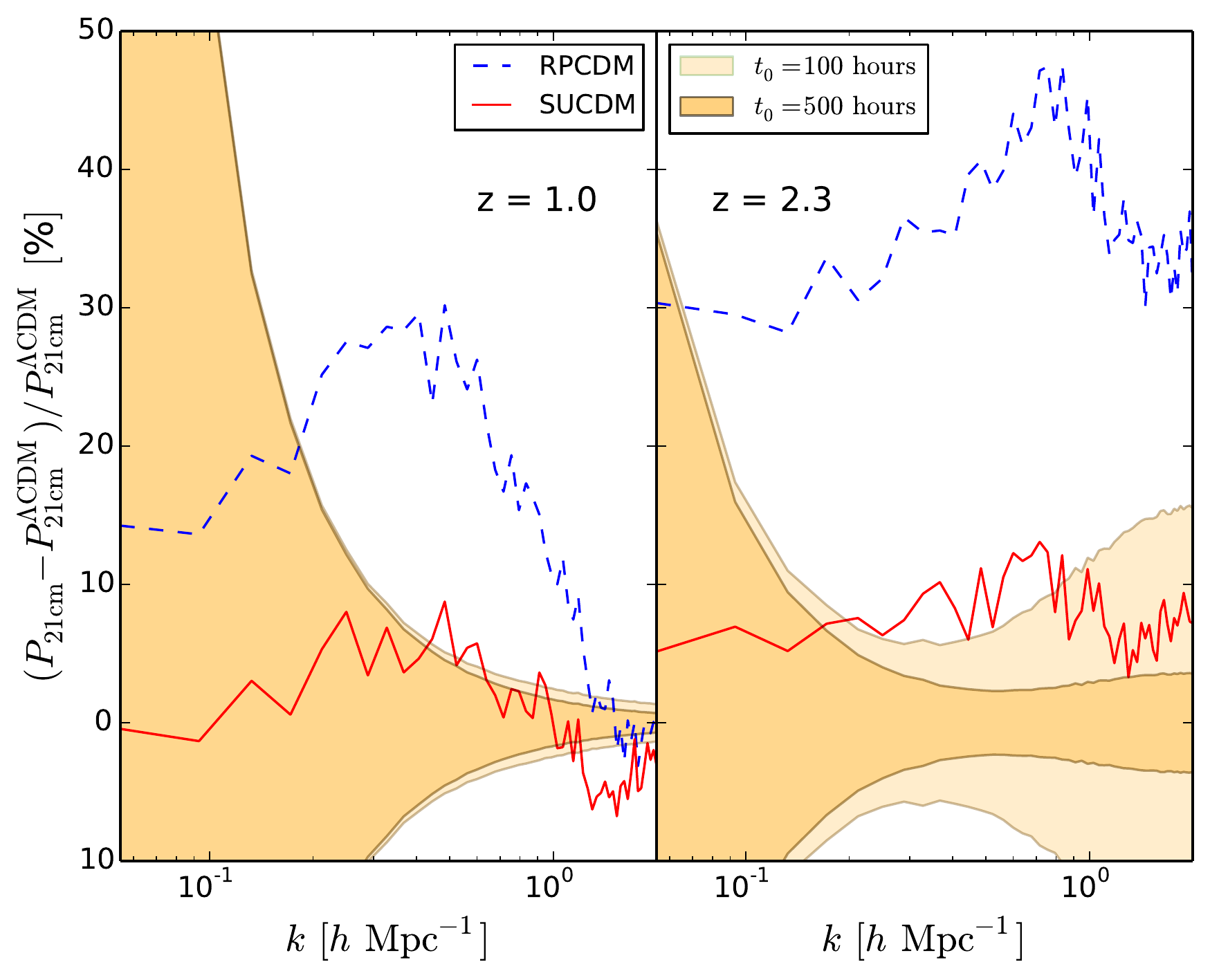}
	\end{minipage}\hspace{0.04\textwidth}%
	\begin{minipage}{0.36\textwidth}
		\caption{\label{SKA_de}  As in Fig.~\ref{SKA_dm} for the non-standard DE models relative to $\Lambda$CDM-W5 model. Figure taken from \cite{carucci_2017}.}
	\end{minipage}
\end{figure}

We use a set of large volume N-body simulations from the Dark Energy Universe Simulations (DEUS) database\footnote{http://www.deus-consortium.org/deus-data/}  of three flat DE models:
\begin{enumerate}
	\item $\Lambda$CDM-W5: a standard cosmological model with cosmological constant $\Lambda$,
	\item RPCDM-W5: a quintessence model with dynamical equation of state as given by the scalar field evolution in a Ratra-Peebles \cite{Ratra1988} self-interacting potential,
	\item SUCDM-W5: a quintessence model with supergravity \cite{SUGRA} self-interacting potential.
\end{enumerate}
The cosmological parameters of all above models have been calibrated in \cite{Alimi2010} in order to reproduce within $1\sigma$ the cosmic microwave background power spectra from WMAP-5 observations \cite{WMAP5} and the luminosity distances from SN-Ia measurements \cite{SN}. In Table~\ref{table1} we show the tuned cosmological parameters  (with $w_0$ and $w_a$ being the DE equation of state parameters of the Linder-Chevalier-Polarski parametrization \cite{Linder,Chevallier} that best-fit the time evolution of the quintessence-field equation of state); the other cosmological parameters are set to $\Omega_b=0.04$, $h=0.72$, $n_s=0.96$ for all models considered. 
All simulations have comoving box size of  $162$ Mpc $h^{-1}$ and follow the evolution of $1024^{3}$ particles. 
We analyse snapshots at $z=1$ and $2.3$.

\begin{table}[h]
		\begin{minipage}{0.60\textwidth}
	\begin{center}
	\lineup
	\begin{tabular}{*{5}{l}}
		\br
		Model & $\Omega_m$ & $\sigma_8$ &  $w_0$ & $w_a$ \\
		\mr
		$\Lambda$CDM-W5& 0.26 & 0.80 & -1 & 0 \\
		RPCDM-W5& 0.23 & 0.66 & -0.87 & 0.08 \\
		SUCDM-W5& 0.25 & 0.73 & -0.94 & 0.19 \\
		\br
	\end{tabular}
\end{center}
	\end{minipage}\hspace{0.04\textwidth}%
	\begin{minipage}{0.36\textwidth}
	\caption{\label{table1} Cosmological model parameters of realistic DE models calibrated against WMAP-5 and SN Ia observations. Details in \cite{Alimi2010}.}
\end{minipage}
\end{table}

As discussed above, the quintessence models investigated are indistinguishable from  $\Lambda$CDM using the considered cosmological data sets. They are characterized by small differences in the large scale clustering that get amplified at small scales as the non-linear regime of gravitational collapse begins \cite{Alimi2010}. Being quantitative, the quintessence models exhibit DM density power spectra in the range $0.1\lesssim k\lesssim 1 \,{\rm Mpc}\,h^{-1}$ and $z\lesssim 2$ of $20-40\%$ lower amplitude than the $\Lambda$CDM prediction. This is caused by the onset of DE dynamics that alters the cosmic expansion during the late matter dominated era by causing a more accelerated expansion at this epoch than in the standard $\Lambda$CDM case. Therefore matter density fluctuations grow less efficiently and the dynamical DE models exhibit a lower level of clustering and, with direct consequences for the HI distribution, lower halo abundances: on average the RPCDM-W5 model has larger deviations ($\sim 20-80\%$) than SUCDM-W5 ($\sim 5-40\%$), as reported in \cite{carucci_2017}, consistently with the linear growth rate of these models. 

In Fig.~\ref{SKA_de} we plot the relative difference of the 21cm power spectrum between the non-standard DE models and the reference $\Lambda$CDM-W5 at $z=1$ (left panel) and $z=2.3$ (right panel). The RPCDM model can be distinguished from the $\Lambda$CDM at high statistical significance. Even the SUCDM model, that features a cosmic expansion and a linear growth rate similar to that of the $\Lambda$CDM, can be potentially distinguished at more that $1\sigma$ at $z=2.3$ in the range of scales corresponding to $0.02\lesssim k\lesssim 2\,{\rm Mpc}^{-1}\,h$.

\medskip
Looking together at Figs.~\ref{WDM}-\ref{SKA_dm}-\ref{SKA_de}, we notice that: $i)$ the 21cm spectra of the DE models differ from the $\Lambda$CDM case not only in amplitude but also in the scale dependence of the signal, especially at small scales where the slope changes; $ii)$ the non-standard DM models differ from the standard cosmological scenario following another trend, with differences increasing at small scales. This suggests that in principle imprints of DE and DM models can be distinguished from one another through 21cm IM observations.

\section{Conclusions}
In this paper I presented some features of the expected 21cm radiation signal in intensity mapping for cosmologies with non-cold DM and non-$\Lambda$ DE. 

I quantified the constraints that actual radio-surveys will be able to set on such models, that are otherwise statistically indistinguishable from standard $\Lambda$CDM with current observational evidences (this holds at all scales for the dynamical DE cases and at large scales for the non-standard DM cases). 

The model dependent features of each scenario are manifest in the 21cm IM spectra such that future SKA measurements should be able to detect or rule out the non-standard DE and DM models we have considered.

\section*{Acknowledgments}
I am happy to thank my collaborators: Matteo Viel, Francisco Villaescusa-Navarro, Pier-Stefano Corasaniti and Andrea Lapi. I also want to thank the students of the Physics Department of the University of Cagliari, who have been amazing hosts of this meeting.

\section*{References} 
\bibliography{bibliography}

\end{document}